\documentclass[12pt,preprint]{aastex}

\begin{document}
\title{Direct Imaging Constraints on the Putative Exoplanet 14 Her c\footnotemark[0]}
\footnotetext[0]{Observations reported here were obtained at the MMT Observatory, a joint facility of the University of Arizona and the Smithsonian Institution.}
\author{Timothy J. Rodigas\altaffilmark{1,2}, Jared R. Males\altaffilmark{1}, Philip M. Hinz\altaffilmark{1}, Eric E. Mamajek\altaffilmark{3}, Russell P. Knox\altaffilmark{1}}

\altaffiltext{1}{Steward Observatory, The University of Arizona, 933 N. Cherry Ave., Tucson, AZ 85721, USA}
\altaffiltext{2}{email: rodigas@as.arizona.edu}
\altaffiltext{3}{University of Rochester, Department of Physics \& Astronomy, Rochester, NY, 14627-0171, USA}

\newcommand{\about}{$\sim$~}
\newcommand{\mj}{M$_{J}$}
\newcommand{\degrees}{$^{\circ}$}
\newcommand{\arcseconds}{$^{\prime \prime}$}
\newcommand{\lprime}{$L^{\prime}$}

\shorttitle{Direct Imaging Constraints on 14 Her c}
\shortauthors{Rodigas et al.}

\begin{abstract}
We present results of deep direct imaging of the radial velocity (RV) planet-host star 14 Her (=GJ 614, HD 145675), obtained in the \lprime ~band with the Clio-2 camera and the MMT adaptive optics system. This star has one confirmed planet and an unconfirmed outer companion, suggested by residuals in the RV data. The orbital parameters of the unconfirmed object are not well constrained since many mass/semimajor axis configurations can fit the available data. The star has been directly imaged several times, but none of the campaigns has ruled out sub-stellar companions. With \about 2.5 hrs of integration, we rule out at 5$\sigma$ confidence $\gtrsim$ 18 \mj ~companions beyond \about 25 AU, based on the \cite{baraffe} COND mass-luminosity models. Combining our detection limits with fits to the RV data and analytic dynamical analysis, we constrain the orbital parameters of 14 Her c to be: $3 \lesssim m/$\mj ~$\lesssim 42$, $7 \lesssim a/$AU $\lesssim 25$, and $e \lesssim 0.5$. A wealth of information can be obtained from RV/direct imaging overlap, especially with deep imaging as this work shows. The collaboration between RV and direct imaging will become more important in the coming years as the phase space probed by each technique converges. Future studies involving RV/imaging overlap should be sure to consider the effects of a potential planet's projected separation, as quoting limits assuming face-on orientation will be misleading.
\end{abstract}

\section{Introduction}
In recent years, several exoplanets have been discovered by direct imaging \citep{marois,kalas,betapic}. While the radial velocity (RV) technique has by far discovered the most exoplanets (525 as of February 2011, http://exoplanet.eu), the few direct imaging discoveries have expanded our knowledge of exoplanets significantly. The directly-imaged planet orbiting Fomalhaut \citep{kalas} shows evidence for having cleared out dusty material in the star's debris disk. The four \about 10 Jupiter-mass (\mj) planets in the HR 8799 system \citep{marois,maroisfourthplanet} orbit at such a wide range of distances that they challenge all current planet formation theories. The sun-like star GJ 758 has a \about 40 \mj ~companion with a T$_{eff} < 600$~K, making it one of the coldest objects ever discovered (\cite{gj758}; \cite{curriegj758}). Recently direct imaging has helped probe exoplanet atmospheres; \cite{spectrum} characterized the atmosphere of HR 8799 c via spectra and \cite{hinz2} characterized the atmospheres of HR 8799 b, c, and d via photometry at 3-5 $\mu$m. For the first time significant orbital motion has been imaged for the planet in the $\beta$ Pic system (\cite{betapic}, \cite{betapicoriginal}). This has helped constrain the planet's orbit.

While the RV technique is mostly sensitive to planets orbiting close (semimajor axis $a \lesssim 5$ AU) to their host stars, direct imaging probes the outer ($a \gtrsim 5$ AU) regions. Thus direct imaging and RV complement each other. A star's planetary system architecture can be fully determined when it is studied by both RV and direct imaging. A planet's true mass, semimajor axis, eccentricity, and orbitial inclination can be determined when it is detected by both RV and direct imaging. Mass-luminosity models for low-mass brown dwarfs and gas-giant planets (e.g, \cite{baraffe,burrows}) can also be constrained and improved in this case. 

The optimal targets for direct imaging are young planets on wide orbits. Since the optimal RV targets are old, quiescent stars with close-in planets, the sample of systems that satisfy both techniques' requirements is currently very small. Even if one cannot currently image any known companions, perhaps imaging of RV stars can detect previously undiscovered companions. \cite{me} used Monte Carlo simulations of RV data to show that as many as 15$\%$ of RV systems that contain a single, moderately-eccentric planet may have an additional massive planet on a wide orbit. In this case the outer planet's RV signal is weak enough that it is not detected above the noise. A more favorable case would be if the outer planet was suggested by a long-term trend in the RV data. Then we know the companion is there, and it is just a matter of detecting it.

\begin{figure}[t]
\centering
\includegraphics[scale=0.5]{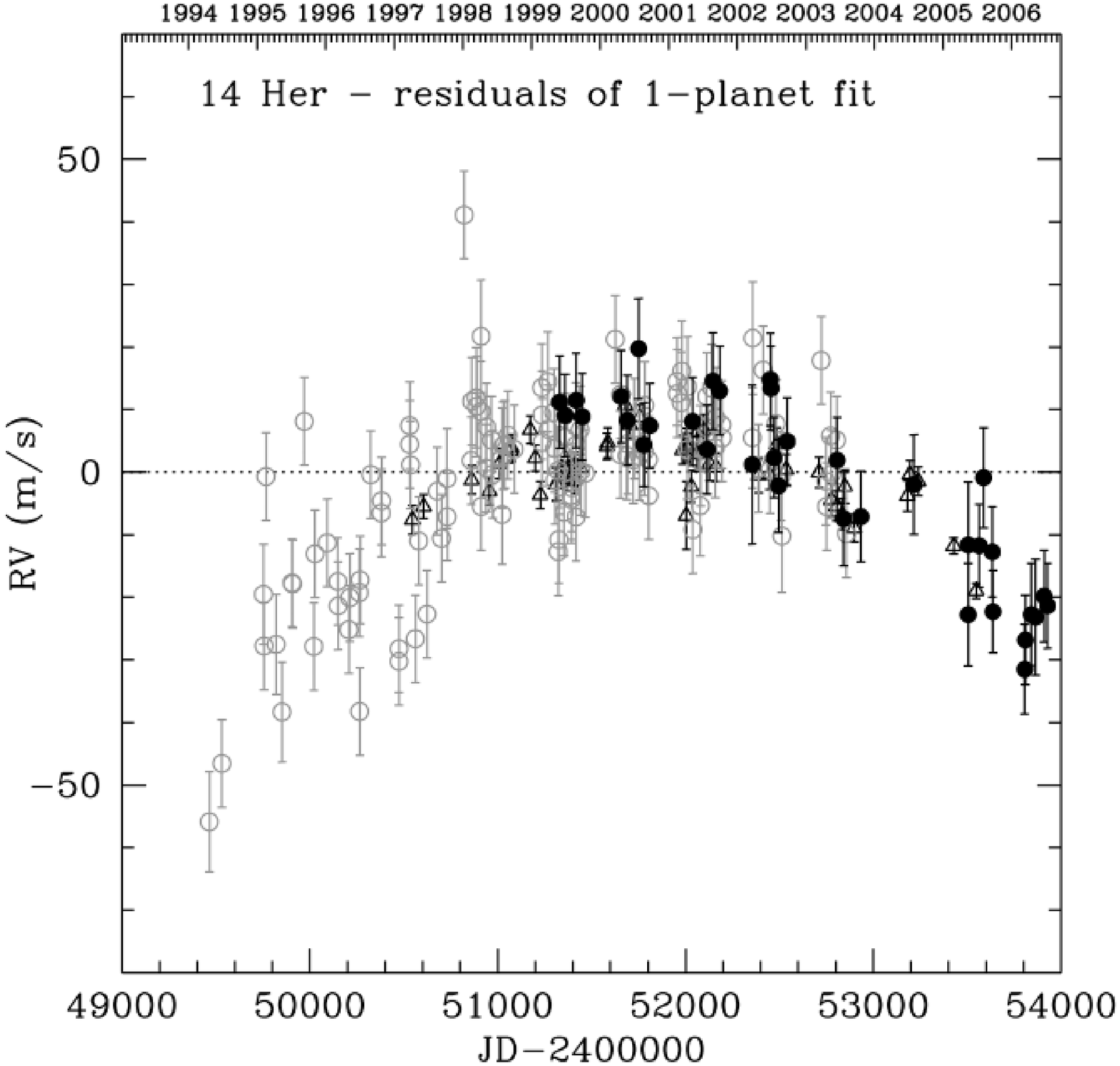}
\caption{RV data excluding the primary (b) component for 14 Her, from \cite{wit}. Because the observation baseline is shorter than the companion's long period, there are many possible good fits to the data and therefore many possible values for 14 Her c's mass, semimajor axis, and eccentricity.}
\label{fig:witdata}
\end{figure}

The star 14 Her (=GJ 614, HD 145675) is a prime target for RV/direct imaging overlap. At a distance of 17.6 pc $\pm$~0.1 pc \citep{updatedhip}, it is close enough that direct imaging can probe as close as \about 9 AU from the star. 14 Her is a multiple-body system. It has one detected planet (of minimum mass 4.64 \mj, semimajor axis $a = 2.77$ AU and eccentricity $e = 0.369$) and a second unconfirmed companion that has been suggested by a long-term trend in the RV data (\cite{wit},\cite{godoriginal},\cite{god}). Two-planet Keplerian fits to the data suggest that the outer companion is $\gtrsim$~2.1 \mj ~and orbits at $\gtrsim$ 7 AU \citep{wit}. Dynamical analysis by \cite{god} suggests a best-fit, minimum $\chi^{2}$ mass of \about 8 \mj ~and a semimajor axis of \about 9 AU, though there were a host of low-$\chi^{2}$ solutions. Fig. \ref{fig:witdata} shows the current RV data excluding the orbit of the primary from \cite{wit}. Due to the lack of data covering the orbit of the outer companion, there are many different possible mass/semimajor axis solutions. 

The companion, at a minimum, orbits on a fairly wide orbit. This is beneficial for direct imaging since luminous objects on wide orbits are easier to detect via imaging than objects on small orbits. This comes from imaging typically having better contrast (and therefore sensitivity) farther from the star. Imaging, especially at the current epoch, is also favorable given that for most of the possible orbital solutions 14 Her c is likely to have a large projected separation; only for some high inclination solutions can the planet be behind or in front of the star.

Direct imaging of stars with known exoplanets has yielded strong constraints on planetary masses orbiting at large separations (e.g., \cite{eind,heinze2,kenworthy}). With regard to 14 Her, there have been several direct imaging campaigns. Two direct imaging campaigns with the Keck and Lick Observatories have already ruled out stellar-mass ($M > 80$ \mj) companions beyond 9 and 12.7 AU, respectively \citep{luhm,patience}.\footnotemark[5] \cite{carson} used the Palomar telescope to rule out \about 70 \mj ~companions beyond \about 18 AU.\footnotemark[5] \footnotetext[5]{These limits assume a face-on orientation to the star, which is unlikely.}Recently \cite{lyot} used the Advanced Electro-Optical System telescope to conduct an imaging survey of nearby solar-like stars, one of which was 14 Her. However with only \about 20 minutes of integration they do not report any significant detection limits for this star. Deeper imaging is required to set meaningful constraints on 14 Her c and any other potential companions.

To investigate the nature of 14 Her c and probe for additional high-mass planets and brown dwarfs, we have carried out deep direct imaging of 14 Her in the \lprime ~band with the MMT adaptive optics (AO) system. In Section 2 we describe the observations and data reduction. In Section 3 we present our contraints on 14 Her c's mass, combining analysis of the published RV data with our direct imaging results and analytic dynamical analysis. In Section 4 we summarize and conclude.

\section{Observations and Data Reduction}
Observations were carried out at the MMT on Mount Hopkins in Arizona on the night of UT 2010 May 30. We used Clio-2 \citep{freed,suresh,hinz,curriegj758}, Arizona's high-contrast near-infrared camera, and observed in the \lprime ~band. The field of view was approximately 9\arcseconds ~by 30\arcseconds , with a plate scale of 29.9 mas/pixel, determined by observations of the binary star HD 223718. We turned the instrument rotator off so that the field of view would rotate throughout the observations; this is essential for high-contrast angular differential imaging (ADI, \cite{adi}). Observing conditions were optimal, with clear skies, good seeing, and the AO providing consistently good atmospheric correction. Throughout the observations, we nodded the telescope every few minutes by 10\arcseconds ~so that each image would contain the target star and sky background. We obtained 8764.7 seconds (2 hrs 26 mins) of integration on 14 Her. At \about 50$\%$ efficiency with Clio-2 in the \lprime ~band, this translated into \about 5 hours of wall-clock time. A small fraction of the obtained images were not used in the data reduction due to the AO being off, wind shake distorting the star, or an infrequent occurrence of losing the star on separate telescope nods. We also obtained unsaturated images of 14 Her and HD 203856 (an L' standard star from \cite{leggett}) using a neutral density filter to calibrate our photometry.

Images were saved and written as stacked data cubes. All data reduction was performed with custom scripts in Matlab. We divided each image by the number of frames in each data cube and by the integration time to produce units of dn/s for each pixel. We corrected the images for bad pixels using a bad pixel mask, and we removed detector and image artifacts as follows: since each image serves as both a target image and a background image, each image was subtracted from the opposite-nod image obtained closest in time to it, as long as that image had not already been used in the nod-subtraction process. Once an image is used to remove sky background, it is no longer available for nod-subtraction. Since we saturated the central 0.\arcseconds 3 core of the star in each image, we determined the pixel location of the star by smoothing each image with a 25 pixel (0.\arcseconds 75) disk (pillbox) average, finding the maximum pixel location, then calculating the center of light at this pixel within a 0.\arcseconds 75 radius. Calculating the center of light allows for sub-pixel registration, which increases contrast. We used the center of light location to register each nod-subtracted image to a common pixel location. We then reduced the images using the LOCI algorithm \citep{loci}. Each image was rotated clockwise by 2.53\degrees ~minus its parallactic angle to obtain north up and east left. The 2.53\degrees ~rotational constant was determined by observing the binary star HD 223718. The final image was produced by median-combining the set of all the images. The change in parallactic angle between the first and last images taken during the night was 158\degrees , allowing us to detect point sources all the way up to the saturated central star (0.\arcseconds 3). 

\section{Results and Discussion}
\begin{figure}[t]
\centering
\includegraphics[scale=1]{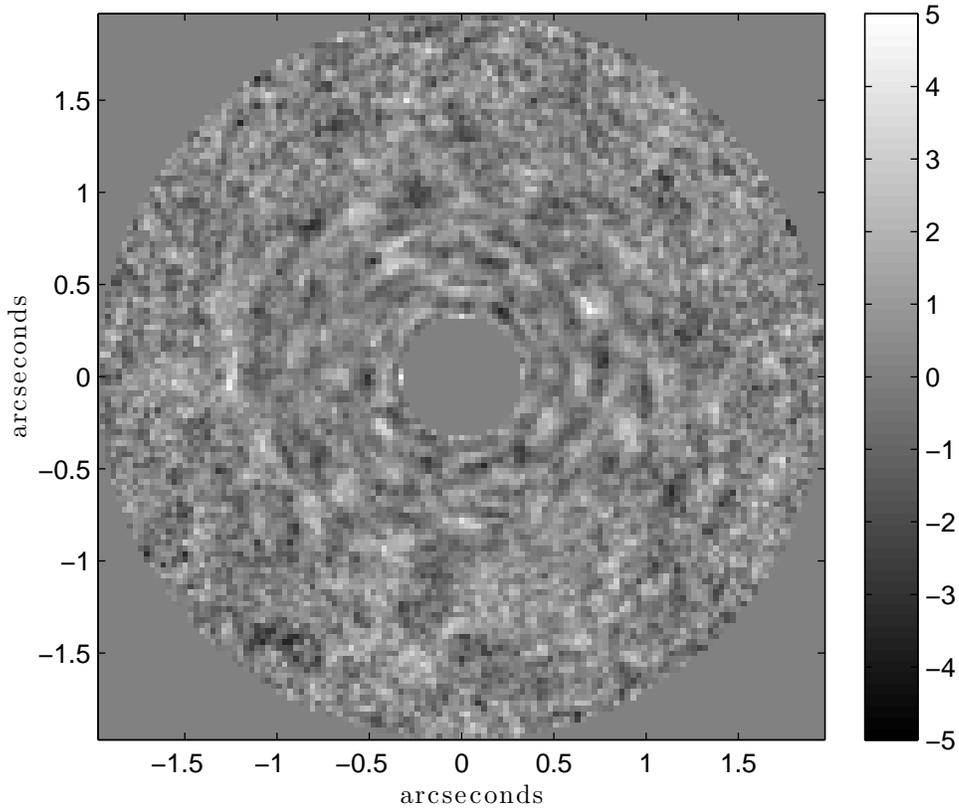}
\caption{Final reduced image of 14 Her, obtained using the LOCI algorithm. North is up and east is to the left. The central 0.\arcseconds 32 has been masked out since this region is saturated. The stretch is -5$\sigma$ to 5$\sigma$, where $\sigma$ was computed as the standard deviation in 5 pixel-wide (0.\arcseconds 15) annuli centered on the star. No candidate companions are identified in the image.}
\label{fig:finalimage}
\end{figure}

Fig. \ref{fig:finalimage} shows our final reduced image of 14 Her. No candidate companions are identified. Some point-sources masquerade as ``real," but after dividing up the data into the first and second half of the night, all sources can be ruled out as speckles.

To determine our sensitivity level, we calculated the standard deviation per pixel in a 5 pixel (0.\arcseconds 15 = FWHM) annulus centered on the star, from 0.\arcseconds 3 to 2\arcseconds . We compared the total flux in a 5-pixel aperture centered on the unsaturated image of HD 203856 to the 14 Her standard deviation per pixel $\times$ the square root of the number of pixels in the 5-pixel aperture. We used this ratio to calculate the apparent magnitude of the 14 Her background as a function of separation from the star. We used unsaturated images of 14 Her to calculate its apparent magnitude relative to the standard star HD 203856. We calculated this value to be 4.76 mag. We used this value and the 14 Her background annulus calculation to determine the contrast as a function of separation from 14 Her. Based on these calculations, we achieve excellent imaging contrast and sensitivity with our observations, reaching contrasts of \about 10 $\Delta$ mags at 0.\arcseconds 4 and $\gtrsim$~13 $\Delta$ mags beyond 1\arcseconds . 

\subsection{Phase Space Constraints}
The true mass of a directly imaged luminous object depends heavily on its age. The younger the host star, the brighter the object is expected to be. Knowing the host star's age accurately is essential for pinning down the companion mass range. Unfortunately 14 Her's age, like many stars' ages, is fairly uncertain. Kinematically the star is a metal-rich thin-disk star, suggesting an age $< 10$ Gyr. The chromospheric activity and age-rotation-activity estimates from \cite{mamajek} estimate an age of 7.8 Gyr and $> 8$ Gyr, respectively. Given that 14 Her is very metal rich ([Fe/H] \about 0.3), age-rotation-activity calibrations (as in \cite{mamajek}) could overestimate the age, since these rely on solar-metallicity calibrator stars. On the other hand, \cite{rochapinto} estimated an age of \about 3.3 Gyr from chromospheric activity and metallicity studies. However this value may underestimate the true age since it is based on a linear age-activity fit, which does not appear to accurately fit samples with well-determined ages \citep{mamajek}. There are in total seven reported age values of 14 Her. One of these is almost twice the age of the universe \citep{takedaapjs}, so we ignore this value. Because our observational result is a non-detection, we want to quote conservative upper limits to mass and should thus use a conservative age in calculations. Therefore we take the maximum of the remaining age values\footnotemark[6] \footnotetext[6]{3.33 Gyr \citep{rochapinto}, 5.0 Gyr \citep{valenti}, 5.24 Gyr \citep{takeda}, 6.9 Gyr \citep{wrightages}, 7.8 Gyr \citep{mamajek}, $> 8$ Gyr \citep{mamajek}} and adopt this value, 8 Gyr, as the age of 14 Her. The COND age tracks do not fall on 8 Gyr, but rather between 5 and 10 Gyr. Therefore we quote \mj ~sensitivities interpolated between these two ages.

\begin{figure}[t]
\centering
\includegraphics[scale=0.85]{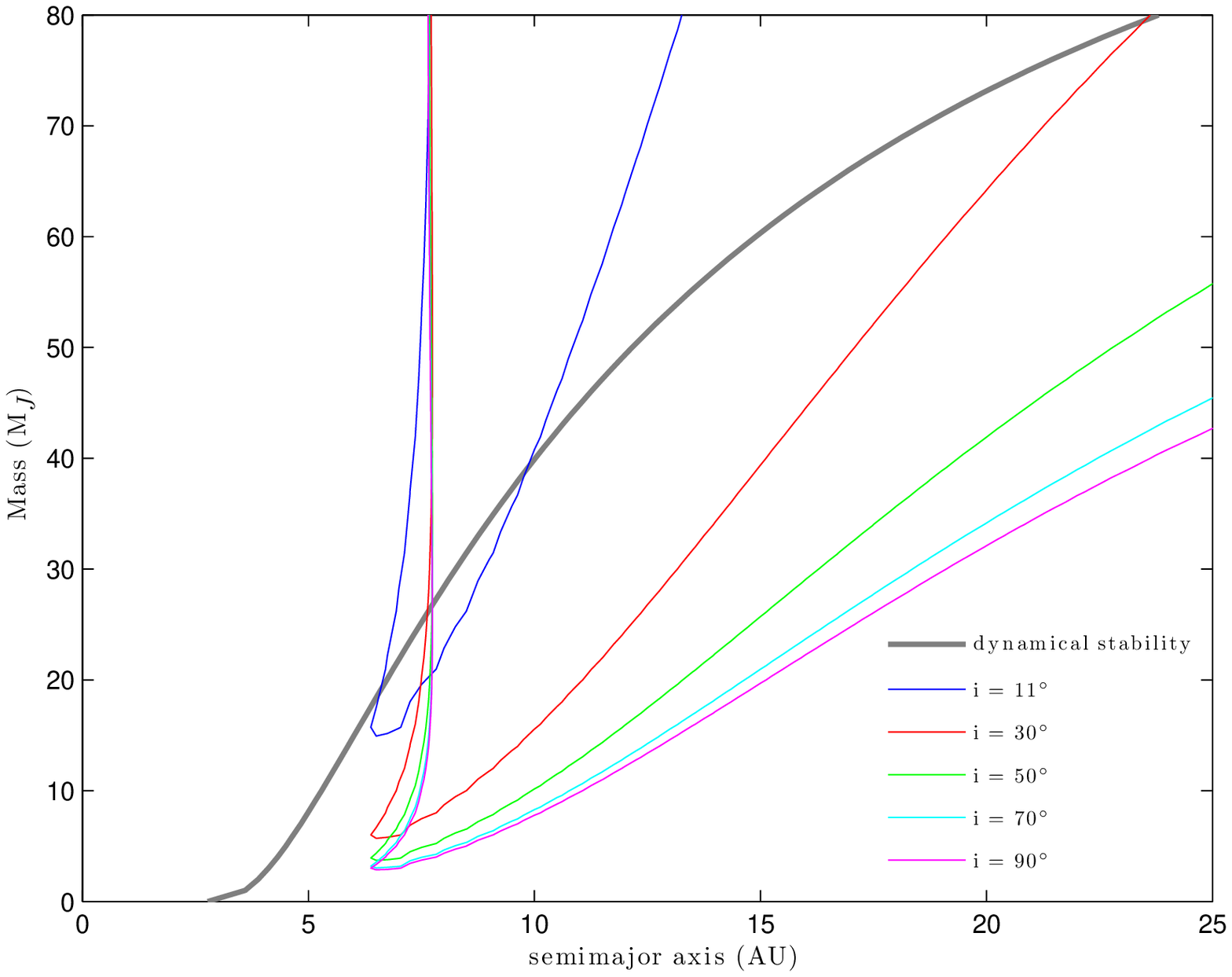}
\caption{Mass vs. semimajor axis for 14 Her c, from $\Delta \chi^{2} = 1$ RV contours (colored lines, $11$\degrees ~$< i < 90$\degrees ) and an analytic dynamical constraint (grey line). The RV contours were computed by fitting the available RV data from \cite{wit}, \cite{butler}, and \cite{naef} at fixed outer planet mass and semimajor axis, allowing all other system parameters to vary. The dynamical constraint, which comes from equations derived by \cite{analyticstability}, represents the maximum mass 14 Her c could have at a given semimajor axis such that the three-body system is stable.}
\label{fig:sens0}
\end{figure}

As a first step towards constraining 14 Her c's phase space, we fit the available RV data from \cite{wit}, \cite{butler}, and \cite{naef} at fixed outer planet mass and semimajor axis, allowing all other system parameters to vary. In Fig. \ref{fig:sens0} we plot the $\Delta \chi^{2} = 1$ contours for inclination angles ranging from 11\degrees ~to 90\degrees ~(colored lines). 14 Her, like most stars with RV planets, has an unknown inclination angle. Nonetheless \cite{han} found $i > 25$\degrees ~and \cite{newastrometry} found $11$\degrees ~$< i < 154$\degrees , both using Hipparcos astrometry. The latter range is equivalent to $11$\degrees ~$< i < 90$\degrees ~since we are dealing with the amplitude of $\sin i$. To be conservative we adopt this range of inclination angles when computing 14 Her c phase space. 

A second constraint comes from analytic dynamical analysis. We used equations derived by \cite{analyticstability} to calculate the maximum mass 14 Her c could have at a given semimajor axis such that the three-body system is stable. This is shown as the grey curve in Fig. \ref{fig:sens0}. The equations assume that both planets have $e = 0$ and treat 14 Her b as a massless test particle. We know that this is not true for 14 Her b; it is both fairly massive ($> 4.64$ \mj ) and eccentric ($e = 0.369$), both of which would limit the outer planet's minimum semimajor axis. This means that our results from this analysis are conservative estimates. Any phase space to the left of this dynamical curve is considered unstable. 

In addition to ruling out mass and semimajor axis values, the dynamical curve also helps constrain eccentricity. We computed eccentricity contours from the RV fits, but do not plot them (for clarity). The $e > 0.5$ solutions are ruled out since the regions of these eccentricity contours that lie within the $\Delta \chi^{2} = 1$ contours correspond to dynamically unstable massive companions. Therefore we take $e = 0.5$ as an upper limit on 14 Her c's eccentricity.

Constraints from our direct imaging cannot simply be plotted in Fig. \ref{fig:sens0} because this would assume a face-on orientation. For imaging we must deal with projected separation instead of semimajor axis. Therefore the most probable orbital elements, which are within the regions to the right of the dynamical curve and within the RV contours, must be mapped into projected separation space at the epoch of the imaging observations. 

\begin{figure}[t]
\centering
\includegraphics[scale=1]{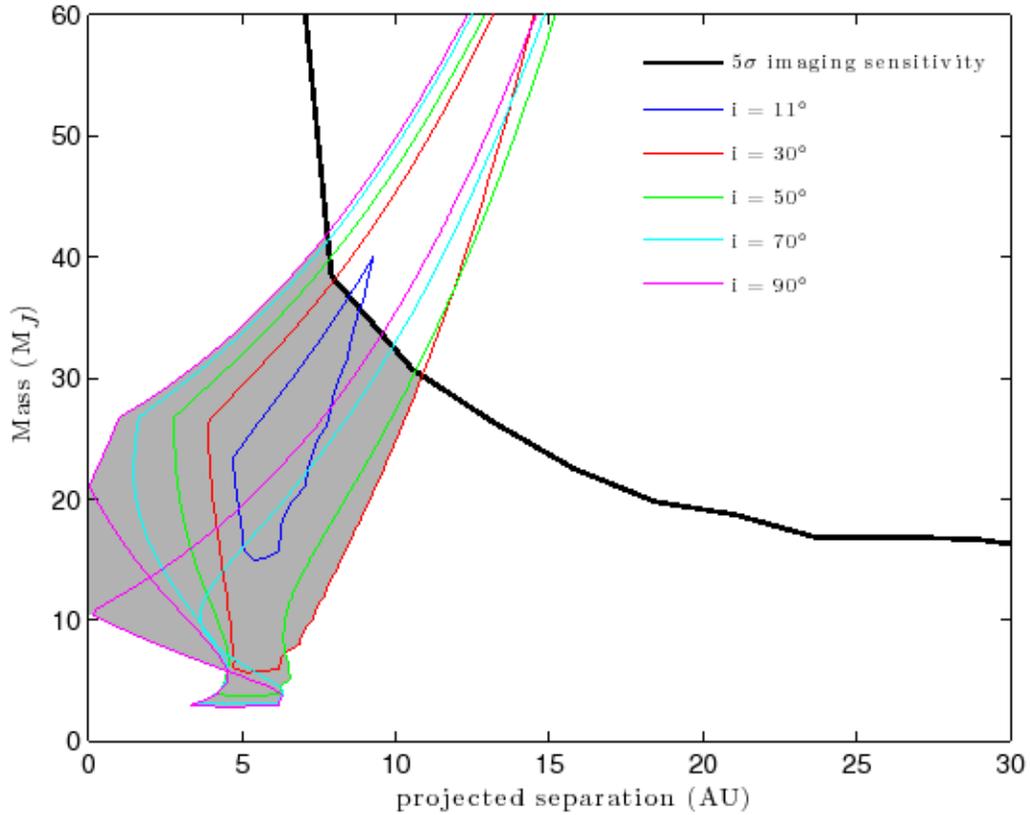}
\caption{14 Her c's mass vs. projected separation at the epoch of our imaging observations. The solid black curve is our imaging sensitivity curve. Any phase space above this line is ruled out at $5\sigma$ confidence. The colored contours represent the projected dynamically-bound $\Delta \chi^{2} = 1$ RV contours. At high inclinations, some orbital solutions place 14 Her c behind or in front of the star due to projection effects. The region shaded grey represents the planet's most probable mass and separation values.}
\label{fig:sensitivity}
\end{figure}

In Fig. \ref{fig:sensitivity} we plot the dynamically-bound RV contours, mapped into projected separation, for $11$\degrees ~$< i < 90$\degrees ~(colored contours). At high inclinations, some orbital solutions place 14 Her c behind or in front of the star due to projection effects. The final constraint comes from our direct imaging sensitivity (solid black curve in Fig. \ref{fig:sensitivity}). The planet's allowed mass and separation values must reside below the imaging curve and inside the dynamically-bound RV contours. We shade this region grey.

\begin{figure}[t]
\centering
\includegraphics[scale=0.85]{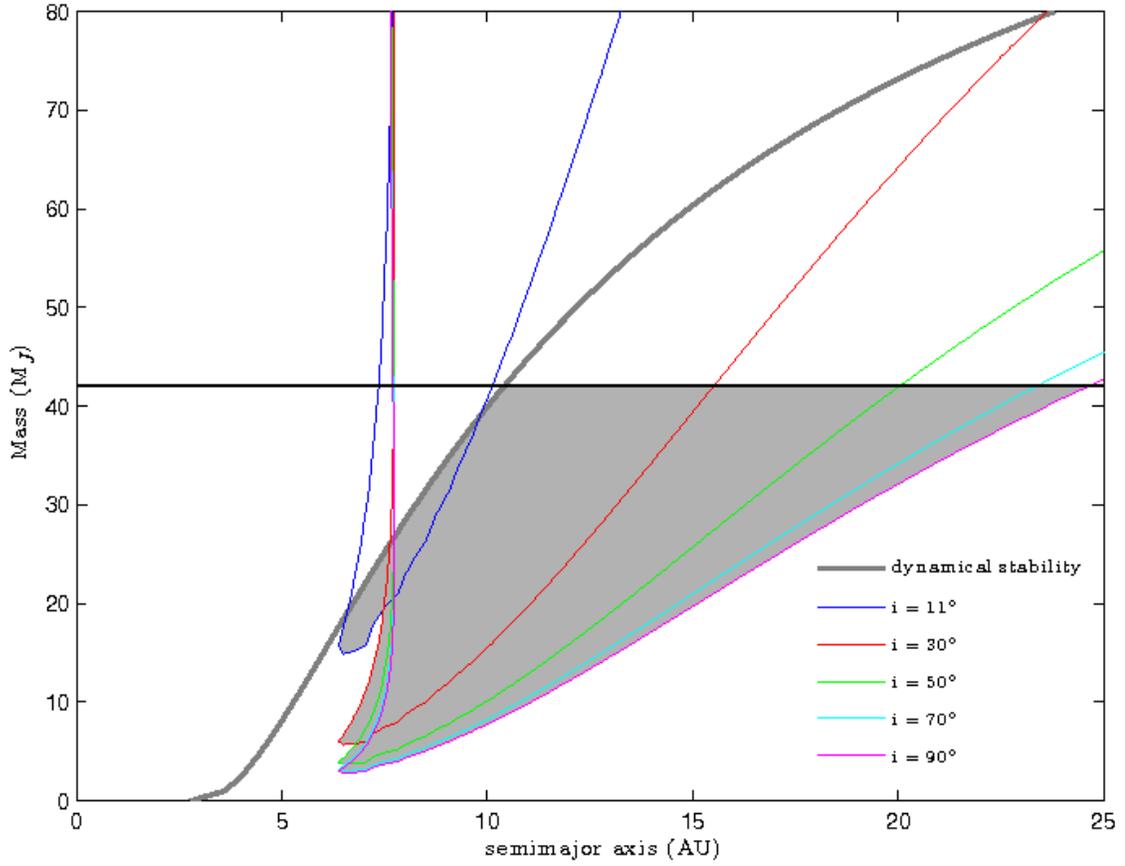}
\caption{The same as Fig. \ref{fig:sens0}, except the 42 \mj ~upper limit on mass (horizontal black line), which was calculated from Fig. \ref{fig:sensitivity}, is included. The phase space above this line is ruled out. The allowed phase space, which is below this line and to the right of the dynamical curve, is shaded grey. 14 Her c's maximum semimajor axis is then \about 25 AU.}
\label{fig:phasespace}
\end{figure}

14 Her c then has the following constraint on mass: $3 \lesssim m/$\mj ~$\lesssim 42$. Using 42 \mj ~as an upper limit on mass, we can constrain 14 Her c's semimajor axis. This is shown in Fig. \ref{fig:phasespace}, which is the same as Fig. \ref{fig:sens0} except that we have included this mass constraint (horizontal black line). The phase space above the horizontal line is ruled out. The allowed phase space, which is below the mass constraint line and to the right of the dynamical curve, is shaded grey. 14 Her c's semimajor axis is then given by $7 \lesssim a/$AU $\lesssim 25$. Our limits agree well with the dynamical results of \cite{god}, who estimated a best-fit mass, semimajor axis, and eccentricity of \about 8 \mj , \about 9 AU, and \about 0.2, respectively.

\section{Conclusions: What is 14 Her c?}
A main goal in exoplanet imaging studies is to determine how planets form. Without better constraints on mass, semimajor axis, and eccentricity we are unable to comment on whether 14 Her c formed by core accretion or disk instability. However, the close separation and low mass constraints presented here indicate the object formed out of the disk, rather than via cloud fragmentation. If 14 Her c's semimajor axis is closer to \about 7 AU, its mass is likely to be small and comparable to that of 14 Her b, which would suggest formation by core accretion. It may take many more years of additional RV and imaging observations before more powerful constraints can shed light on how the two planets formed. Nonetheless this work is important for showing the potential of RV/direct imaging overlap. We have demonstrated the ``proof of concept," showing how much orbital phase space can be constrained with just a few hours of observations combined with RV data and simple dynamical analysis. We also showed the importance of considering projected separation of an RV planet when constraining phase space. Future cases of RV/imaging overlap should work in projected separation, as assuming face-on orientations can be misleading. The next generation of large ground- and space-based telescopes will be able to probe lower-mass regimes and will therefore set more powerful constraints. This deeper imaging, combined with continued RV monitoring of systems like 14 Her, will help characterize the architectures of the planetary systems beyond our own. 

\acknowledgements
We thank K. Goździewski and colleagues for sharing their dynamical analysis data. We thank Mike Alegria for operating the telescope and the AO system so smoothly during the observations. We thank the UAO telescope allocation committee for generously providing telescope time for these observations. We thank Eric Nielsen for allowing us to use his orbital element calculator code. We thank the anonymous referee for a very thorough and quickly-received referee report.


\end{document}